\documentclass[aps]{revtex4}
\usepackage{amscd}
\usepackage{amssymb}
\usepackage{amsbsy}
\usepackage{amsthm}
\usepackage{amsmath}

\newcommand{\proofend}{\hfill\rule{0.2cm}{0.2cm}}

\theoremstyle{plain}
\newtheorem{theorem}{Theorem}[section]

\newtheorem{proposition}[theorem]{Proposition}
\newtheorem{definition}[theorem]{Definition}

\theoremstyle{remark}

\newtheorem{assumption}{Assumption}

\begin{document}

\title{(In)finiteness of Spherically Symmetric Static Perfect Fluids}
\author{J. Mark Heinzle}
\email{mheinzle@galileo.thp.univie.ac.at}
\affiliation{Institute for Theoretical Physics, University of Vienna, 
Boltzmanngasse 5, A-1090 Vienna, Austria\\}

\begin{abstract}
\noindent
This work is concerned with the finiteness problem for static, 
spherically symmetric perfect fluids
in both Newtonian Gravity and General Relativity.
We derive criteria on the barotropic equation of state
guaranteeing that the corresponding perfect fluid solutions possess
finite/infinite extent.
In the Newtonian case, for the large class of monotonic equations of state, 
and in General Relativity we improve earlier results. 
Moreover, we are able to treat the two cases in a completely
parallel manner, which is accomplished by
using a relativistic version of Pohozaev's identity
in the proof of the relativistic criterion. 
This identity and further generalizations are presented
in detail.

%

\vskip1cm
\noindent
{\bf Physics and Astronomy Classification Scheme (2001):} 
Primary: 04.40.Nr; 
secondary: 04.40.Dg, 
02.40.Hw, 
02.40.Ky. 
\end{abstract}
\maketitle

\section{Introduction}

In this paper we consider non-rotating stellar models, i.e.,
we consider static, self-gravitating perfect fluids
with a barotropic equation of state $\rho(p)$ relating density and
pressure. The basic equations are the Euler-Poisson equations
in Newtonian theory and the Euler-Einstein equations in General Relativity.
We are concerned with globally regular solutions on $\mathbb{R}^3$ 
consisting of a perfect fluid region and possibly a vacuum region.
We focus on spherically symmetric solutions in particular.

In Newtonian theory spherical symmetry is no restriction; 
static stellar models are necessarily spherically symmetric
(see \cite{Lindblom:1993} for an overview).
In General Relativity, although conjecturable,
a general theorem establishing spherical symmetry of
solutions does not exist so far, although some progress
has been made \cite{Beig/Simon:1991, Beig/Simon:1992}
\cite{Lindblom/Masood-ul-Alam:1994}.
Some more remarks follow below.

Existence and uniqueness of spherically symmetric solutions
has been proven satisfactorily in Newtonian theory
(for rather general equations of state \cite{Schaudt:2000})
as well as in General Relativity (for smooth equations of state \cite{Rendall/Schmidt:1991}).
For results regarding topology see \cite{Masood-ul-Alam:1987}.

Mathematical issues apart, perfect fluid solutions
are of interest in astrophysics. Primarily, perfect fluid
solutions represent stellar models (we refer to the classic
\cite{Chandrasekhar:1957}; a recent review addressing some issues
is \cite{Beig/Schmidt:2000}). 
But also other astrophysical objects as globular clusters
possess descriptions in terms of perfect fluid solutions \cite{Spitzer:1987}.

The main question we pose in this work is the (in)finiteness question.
Under which conditions on the equation of state does the
corresponding perfect fluid solution possess finite or infinite extent?
We briefly recall some criteria existing in the literature:
The polytropic equations of state $p = K \rho^{(n+1)/n}$ ($K>0$, $n>0$ constants)
have been studied extensively, both analytically and numerically.
In Newtonian theory finiteness is guaranteed for $n<5$, $K$ arbitrary 
(for a general discussion see \cite{Horedt:1987}), in General Relativity
the case is considerably more complex \cite{Nilsson/Uggla:2000}.
Criteria for more general classes of equations of state 
are, e.g., the following:
In Newtonian theory and in General Relativity,
spherical symmetry presupposed, the perfect fluid solution is
finite, if $\rho|_{p=0} > 0$.
This result can be subsumed under the criterion guaranteeing finiteness
of the fluid configuration if
$\int_0^p dp^\prime \rho(p^\prime)^{-2} < \infty$ holds.
In the case $\rho|_{p=0}=0$ monotonicity 
of the equation of state must be assumed here.
As a counterpart to this criterion, under the same assumptions,
there exists the following theorem:
If $\int_0^p dp^\prime \rho^{-1}(p^\prime) \not< \infty$ (Newtonian theory)
or $\int_0^p dp^\prime (\rho(p^\prime)+c^{-2} p^\prime)^{-1} \not< \infty$
(General Relativity), 
then the fluid solution must necessarily extend to infinity
(see \cite{Rendall/Schmidt:1991} for a good presentation of these criteria).
Note incidentally that the last two quantities will play an important role for our
considerations as well (see below).
For equations of state of the type 
$p = K \rho^{(n+1)/n} ( 1 + O(\rho^{1/n}))$ (as $\rho\rightarrow 0$)
with $1<n<3$ finiteness of the fluid solution has be proven in \cite{Makino:1998}
(some detail follow below). Note that these criteria
involve the behavior of the equation of state for small $p$ only.
More general criteria, however, must be based on the behavior of
the equation of state $\rho(p)$ for all $p$. 
To show that consider the polytropic equation of state for $n=5$ in
Newtonian theory as an example. Perturbing $\rho(p)$ in
a small neighborhood of some finite value of $p$ suffices
to produce either finite or infinite fluid configurations.
Criteria which are capable of dealing with such phenomena
have been derived in \cite{Simon:1993} and \cite{Simon:2001}.
Some of those criteria will be reproduced in this paper
(see, e.g., theorems \ref{J0theorem} and \ref{relI0theorem});
the other criteria we present here can be regarded
as generalizations or modifications in a certain sense.
In \cite{Simon:2001} the assumptions on the equation
of state and the solutions are kept rather general,
several theorems are formulated for Sobolev functions.
Occasionally we give cross references to \cite{Simon:2001}.

The paper is divided into three main parts.
Part one (Newton):
In sections \ref{newton:basics}--\ref{newton:pohozaevandcriteria} we treat
the (in)finiteness question for perfect fluid solutions in Newtonian theory.
Section \ref{newton:basics} is concerned with the Euler-Poisson equations;
in section \ref{criteria} some quantities are studied which are 
necessary to formulate the (in)finiteness criteria in section \ref{newton:pohozaevandcriteria}
(theorem \ref{J0theorem} and theorem \ref{J-1theorem}).
The crucial tool for the proofs of the theorems is Pohozaev's identity
\cite{Pohozaev:1965}.
Part two (General Relativity):
Sections \ref{euler-einstein}--\ref{einstein:pohozaevandcriterion}
deal with general relativistic perfect fluid solutions.
The presentation parallels the Newtonian case in order to facilitate comparison;
section \ref{euler-einstein}: the Euler-Einstein equations, section \ref{einstein:j0}:
definitions.
The (in)finiteness theorems are formulated in section \ref{einstein:pohozaevandcriterion}; 
the main ingredient for the proofs is a relativistic generalization
of Pohozaev's identity.
To conclude the first two parts of the paper, 
in section \ref{examples} we discuss applications of the criteria we have derived
both for the Newtonian and the relativistic case.
Part three (Pohozaev-like identities):
In section \ref{methodofconstruction} we present a powerful
method of deriving ``Pohozaev-like'' identities. In particular
we treat the relativistic Pohozaev identity which has been used
for the proof of theorem \ref{relJ0theorem} in section \ref{einstein:pohozaevandcriterion}.

\section{Newton: Basics}
\label{newton:basics}

Newtonian static perfect fluids are regular solutions to the Euler-Poisson equations
(on $(\mathbb{R}^3, \delta_{ij})\,$), given
a barotropic equation of state $\rho(p)$ relating the pressure and the density. 
\begin{subequations}\label{poissoneuler}
\begin{align}\label{poisson}
\Delta u(x) &= 4\pi \rho(x) \\ \label{euler}
\partial_i p(x) &= -\rho \partial_i u(x)
\end{align}
\end{subequations} 
Here, $u$ denotes the Newtonian potential, $\partial_i = \frac{\partial}{\partial x^i}$
and $\Delta = \partial^i \partial_i$. 

In section~\ref{criteria} we discuss 
which classes of equations of state we consider
in this paper (see definition~\ref{Gammadef} ff.).
In all cases under consideration 
the potential $u$ can be viewed as a function of $p$; integrating~(\ref{euler}) we obtain  
\begin{equation}\label{up-us} 
u(p) - u_S = -\int_0^p dp^\prime \rho^{-1}(p^\prime)
=: -\Gamma(p)
\qquad\mbox{where}\quad u_S := u|_{p=0} \:.
\end{equation}
We distinguish solutions of (\ref{poissoneuler})
according to whether they extend to infinity or not.
A solution with finite extent (surface $\{p=0\}$) possesses the 
surface potential $u_S$. At the surface the interior solution
is joined to an exterior vacuum solution (standard junction conditions).
Solutions extending to infinity satisfy $p>0$ everywhere.

We define the {\it ``normalized''} potential $v$ as $v := u - u_S$.
Obviously, $v(p)$ is monotonic with $v|_{p=0} = 0$. Consequently, both $p$ and $\rho$
can be viewed as functions of $v$, and, moreover,
$p(v) = -\int_0^v \rho(v^\prime) dv^\prime$. 
For monotonic equations of state $\rho(p)$, $\rho(v)$ is monotonic as well
(the converse also being true).

{\it Remark}.
The quantities $p(x), \rho(x)$ vanish at infinity, i.e.,
$p(x),\rho(x) \rightarrow 0$ ($\|x\| \rightarrow \infty$) 
for solutions of (\ref{poissoneuler}).
In spherical symmetry this can be proven rather easily (see, e.g., \cite{Berestycki:1981}).
Using the standard asymptotic condition for $u(x)$, it follows that
$u(x)\rightarrow 0$ ($\|x\| \rightarrow \infty$).
As a consequence, $v = u$, if the fluid has infinite extent. 

{\it Remark}.
Newtonian static perfect fluid solutions are necessarily spherically symmetric.
For a short overview on the topic of symmetries of stellar models 
see \cite{Lindblom:1993} (and references therein).

{\it Remark}. 
The potential $u(r)$ ($r=\|x\|$) is monotonically increasing. This
follows directly from the maximum principle applied to (\ref{poisson}).
Hence, $u(r)$ assumes its minimal value $u_c$ at the center.
Correspondingly, $p(r)$ is a strictly decreasing function (with maximum $p_c$),
which results from monotonicity of $\Gamma(p)$ in (\ref{up-us}).
$p_c$ ($u_c,\ldots$) is a suitable parameter to characterize the one-parameter
family of regular solutions to~(\ref{poissoneuler}).

\section{Newton: The quantities $I_i$}
\label{criteria} 

Throughout this paper we consider barotropic equations of state $\rho(p)$
given at least on an interval $[0,p_{max}]$. We assume that $\rho$
is positive for all $p \in [0,p_{max}]$.
Recall that perfect fluid configurations are always finitely extended 
in the case $\rho|_{p=0} > 0$. Therefore 
we focus on equations of state with $\rho|_{p=0}=0$.
``Microscopically stable''
matter (see, e.g., \cite{Harrison/Thorne/Wakano/Wheeler:1983}) is described by monotonic
equations of state, i.e., by increasing functions $\rho(p)$. Often we will
restrict ourselves to this class of equations of state.

To formulate our theorems in section \ref{newton:pohozaevandcriteria} we now
introduce some quantities. Note that all
these definitions depend on the equation of state only.

\begin{definition}\label{Gammadef}
Consider a barotropic equation of state $\rho(p)$ as described above with
$\rho(p)$ piecewise ${\mathcal C}^0$ on $[0, p_{max}]$.
Let $\Gamma(p)$ be defined as
\begin{equation}\label{Gamma}
\Gamma(p) := \int_0^p dp^\prime\,\rho^{-1}(p^\prime) \:.
\end{equation}
\end{definition}

\begin{assumption}\label{gammaexists}
Throughout this paper we  
assume that
$\Gamma(p)$ exists for some $p>0$. 
Obviously, $\Gamma(p)$ is ${\mathcal C}^0[0,p_{max}]$.
\end{assumption}

\begin{assumption}\label{limitrho-1pexists} 
We require the limit $\lim_{p\rightarrow 0} \rho^{-1} p$ to exist.
\end{assumption} 

{\it Remark}.
Note that assumptions~\ref{gammaexists} and~\ref{limitrho-1pexists}
are independent. However,
if the equation of state $\rho(p)$ is monotonic (at least on $[0,\epsilon]$ for some $\epsilon>0$), 
then assumption~\ref{limitrho-1pexists} follows from assumption~\ref{gammaexists}.
Note further that $\lim_{p\rightarrow 0}\rho^{-1} p = 0$, if \ref{gammaexists} 
and~\ref{limitrho-1pexists} hold.
For the details we refer the reader to appendix \ref{appendixA}.


{\it Remark}.
Provided that $\rho(p)$ is at least piecewise ${\mathcal C}^1$ we may investigate
$\frac{d\rho}{dp}$.
From $\lim_{p\rightarrow 0}\rho^{-1} p = 0$ we conclude 
$\lim_{p\rightarrow 0} (\frac{d\rho}{dp})^{-1} = 0\,$, if this limit exists.
For details see appendix \ref{appendixA}.

\begin{definition}\label{js} 
Consider an equation of state $\rho(p)$ satisfying 
assumptions~\ref{gammaexists} and~\ref{limitrho-1pexists}. We define the following
quantities:
\begin{subequations}
\begin{align}\label{J-1}
I_{-1}(p) & := 7 \int_0^p dp^\prime\, \Gamma(p^\prime) - 6\, \Gamma(p) p \\\label{J0} 
I_0(p) & := \Gamma(p) - 6 \rho^{-1} p \\ \label{J1}
I_1(p) & := 6 p \,\rho^{-2}\, \frac{d\rho}{dp} - 5 \rho^{-1} \\ \label{I2}
I_2(p) & := 5 p \,\rho^{-2} \,\frac{d^2\rho}{dp^2} + p \,\rho^{-3} (\frac{d\rho}{dp})^2
\end{align}
\end{subequations}
We require $\rho(p)$ to be piecewise ${\mathcal C}^0$ on $[0,p_{max}]$; in order
to define $I_1$ let $\rho(p)$ be ${\mathcal C}^0[0,p_{max}]$ and piecewise
${\mathcal C}^1$; for $I_2$ we require differentiability of one order higher.
\end{definition}

{\it Remark}.
From the comments above we conclude that 
$I_0(p)$ is well-defined and piecewise continuous on $[0,p_{max}]$.
Moreover, $I_0|_{p=0} = 0$. 

\begin{proposition}\label{Jinclusions}
Let $\{I_i \leq 0\}$, $\{I_i \equiv 0\}$, and $\{I_i \geq 0\}$ denote
the sets of all equations of state $\rho(p)$ 
(as in definition~\ref{js}) such that, for all $p \leq p_{max}$,
$I_i \leq 0$, $I_i \equiv 0$, or $I_i \geq 0$ respectively.
Then
\begin{subequations}
\begin{align}
\{I_1 \leq 0\} &\subset \{I_0 \leq 0\} \subset \{I_{-1} \leq 0\} \\
\{I_1 \equiv 0\} &\equiv \{I_0 \equiv 0\}\equiv \{I_{-1} \equiv 0\}\\
\{I_1 \geq 0\} &\subset \{I_0 \geq 0\} \subset \{I_{-1} \geq 0\} \:.
\end{align}
\end{subequations}
\end{proposition}

{\it Proof.}
$I_i$ is the integral of $I_{i+1}$ ($i=-1,0$), i.e.\
\begin{equation}
\frac{d}{dp}\, I_{-1}(p) = I_0(p) \qquad
\frac{d}{dp}\, I_0(p) = I_1(p)\:.
\end{equation}
Together with $I_0|_{p=0} = 0$ and $I_{-1}|_{p=0} = 0$, the claim is proven.\proofend

{\it Remark}.
$\{I_i\equiv 0\}$ ($i=-1,0,1$) corresponds to the set of polytropic equations of state 
$p = K\,\rho^{\frac{n+1}{n}}$ ($K=\mbox{const}$) with
polytropic index $n=5$. This can be obtained easily by solving the differential
equation given by $I_1\equiv 0$.

{\it Remark}.
We also have the relations $\{I_2 \equiv 0\} \supset \{I_1 \equiv 0\}$ and
$\{I_2 \leq 0\} \subset \{I_1\leq 0\}$. For a proof see \cite{Beig/Simon:1991,Beig/Simon:1992}.

{\it Comments}.
In this paper we will only be concerned with the quantities $I_0$ and $I_{-1}$.
In \cite{Simon:2001} it has be shown that 
$I_1$ plays the main role in a theorem related to part A of theorem~\ref{J0theorem} 
and~\ref{J-1theorem}\,; basically it states that by 
controlling the sign of $I_1$ it becomes possible
to get rid of the AFMD requirement
(see section~\ref{newton:pohozaevandcriteria} for the context).
$I_2$ itself is not as relevant as its relativistic counterpart; some
comments follow in section~\ref{einstein:j0}.

\section{Newton: Criteria}
\label{newton:pohozaevandcriteria} 

In the present section we prove the main theorems
in the Newtonian case (theorems \ref{J0theorem} and \ref{J-1theorem}).
These theorems formulate criteria on the equation of state
ensuring (in)finiteness of the corresponding static perfect fluid solutions.
The main ingredient for the proofs is Pohozaev's identity \cite{Pohozaev:1965}
(proposition \ref{pohoprop}). 

{\it Remark}. 
For the following we assume that the equation of state $\rho(p)$ 
(and thus $\rho(v)$) is at least piecewise ${\mathcal C}^0$,
that the function $p(r)$ is continuous and piecewise ${\mathcal C}^1$,
$\rho(r)$ piecewise ${\mathcal C}^0$, and $u(r)$ ${\mathcal C}^1$ and
piecewise ${\mathcal C}^2$. 
These requirements are sensible: In~\cite{Schaudt:2000}, beyond 
existence and uniqueness, it has been shown 
that for the large class of so-called admissible equations of state 
(including step functions, polytropic behavior, etc.) 
the (in general non-classical) solutions $u(r)$ of (\ref{poissoneuler}) are 
${\mathcal C}^1$.
If in addition the equation of state is piecewise
${\mathcal C}^{k,\alpha}$ (with $k\geq 0, 0<\alpha<1$),
then $u(r)$ is piecewise ${\mathcal C}^{k+2,\alpha}$ (and
therefore ${\mathcal C}^{k+2}$).
Here, ${\mathcal C}^{k,\alpha}$ means that the $k$th derivative
is H\"older continuous. 

\begin{definition} \cite{Beig:2000}. \label{sigmadefi} 
From $v$, its derivatives, and from $p$, $\rho$ we define
the following symmetric tensor $\sigma_{ij}$ on $\mathbb{R}^3\,$:
\begin{equation}\label{sigmaij} 
\sigma_{ij} := -2 v v_{,ij} + 6 v_{,i} v_{,j} - 2 \delta_{ij} v_{,k} v^{,k} +
8 \pi \delta_{ij} ( \rho v + 4 p)
\end{equation}
\end{definition}

$\sigma_{ij}$ is divergence free, $\partial^j \sigma_{ij} = 0$; its trace
$\sigma^i_i = 16 \pi (\rho v + 6 p)$.
Note that by (\ref{poissoneuler}) $v_{ij}$ can be written as first derivatives.
Pohozaev's identity 
is a direct consequence.
In its present form it is due to \cite{Beig:2000, Simon:1993}.

\begin{proposition}(Pohozaev identity).\label{pohoprop}
With $\xi^j$ a dilation, i.e., $\xi^j = x^j$ we have the following identity:
\begin{equation}\label{pohozaev}
\partial^i(\sigma_{ij} \xi^j) = 16\pi (\rho v + 6 p )
\end{equation}
\end{proposition}

\begin{definition}(Asymptotic conditions).
A solution of (\ref{poissoneuler}) is called AFMD (asymptotically flat with
mass decay conditions), if for some $\epsilon>0$
\begin{equation}\label{NewtonAMFD}
u = O^\infty(\|x\|^{-1}) \qquad
\rho = O^\infty(\|x\|^{-3-\epsilon})\:.
\end{equation} 
Consequently, using Euler's equation~(\ref{euler}), $p= O^\infty(\|x\|^{-4-\epsilon})$.
For a generalization in terms of Sobolev spaces see \cite{Simon:2001}.
\end{definition} 

\begin{proposition}(Pohozaev integrated).\label{poho}
Consider a static perfect fluid solution in Newtonian theory,
i.e., let $v(r) = u(r) - u_S$, $p(r)$, $\rho(r)$ be a solution of
(\ref{poissoneuler}). Furthermore, assume AFMD.
Then, 
\begin{subequations}\label{pohozaevintegrated}
\begin{align}\label{finitepohozaev}
\int\limits_{\mathbb{R}^3} d^3x \,(\rho v + 6 p) &= \frac{M^2}{R}\:, 
&& \mbox{for solutions with finite extent, and,}
\\[0.1cm]\label{infinitepohozaev} 
\int\limits_{\mathbb{R}^3} d^3x \,(\rho v + 6 p) &= 0 \:,
&& \mbox{if the fluid extends to infinity.}
\end{align}
\end{subequations}
In (\ref{finitepohozaev}), $M$ denotes the mass, and $R$ the radius of the finite
fluid object.
\end{proposition}

{\it Proof.}
Integrating (\ref{pohozaev}) for fluid solutions with finite extent 
we get for the l.h.\ side
\begin{equation}
\int\limits_{\mathrm{Ball}(R)} d^3x \partial^i (\sigma_{ij} \xi^j) = 
4\pi R^2 \sigma_{ij}\big|_R\, x^j R^{-1} x^i = 16 \pi R^3 v_{,r}^2\big|_R\:.
\end{equation}
The exterior solution is vacuum, i.e., $u = -M/r$ ($r\geq R$), so that
$v_{,r}|_R = u_{,r}|_R = M R^{-2}$. 
For solutions extending to infinity, 
$\int_{\mathrm{Ball}(r)}d^3x \partial^i (\sigma_{ij} \xi^j) = 
4\pi r \sigma_{ij} x^j x^i$. Making use of the AFMD
conditions, this expression can be shown to converge to zero as $r\rightarrow\infty$. \proofend

{\it Remark.}
The identities (\ref{finitepohozaev}) plus (\ref{infinitepohozaev})
can be combined to give
\begin{equation}\label{bothpohozaev} 
\int\limits_{\mathbb{R}^3} d^3x \,(\rho u + 6 p) = 0\:.
\end{equation}
This is because, for the finite body case, 
$\int d^3x \rho u = \int d^3x \rho (v + u_S) = \int d^3x \rho v + u_S M$,
and $u_S = -M/R$.

\begin{theorem}\cite{Simon:1993,Simon:2001}.\label{J0theorem} 
Consider an equation of state $\rho(p)$ being piecewise ${\mathcal C}^0$
and satisfying assumptions~\ref{gammaexists} and~\ref{limitrho-1pexists}. 
\begin{enumerate}
\item[{\bf A}] Let $(v=u-u_S, \rho ,p)$ be an AFMD solution to the Euler-Poisson 
equations~(\ref{poissoneuler}) with $p\leq \tilde{p}$ for some $\tilde{p}>0$.
If $I_0 \leq 0$ ($I_0 \not\equiv 0$) for all $p\in [0,\tilde{p}]$, 
then the solution has finite extent.
\item[{\bf B}] If $I_0 \geq 0$ ($I_0 \not\equiv 0$) for all $p\in [0,\tilde{p}]$,
then there is no AFMD solution to the Euler-Poisson 
equations~(\ref{poissoneuler}) satisfying $p\leq \tilde{p}$.
\end{enumerate}
\end{theorem}

{\it Proof.}
$I_0(p) = \Gamma(p) - 6 \rho^{-1} p = - (v(p) + 6 \rho^{-1} p)$. Hence, the
l.h.\ side of~(\ref{pohozaevintegrated}) equals
$-\int d^3x \rho I_0(p(x))$. 
The rest follows easily from proposition~\ref{poho}.\proofend

{\it Remark}. Since $p(r)$ is a monotonically decreasing function,
$p\leq \tilde{p}$ corresponds to $p\leq p_c$, where $p_c$ is the
central pressure of the perfect fluid solution.

{\it Remark}. In the proof only the quantity $\rho I_0$ was used. Thus,
proposition~\ref{J0theorem} also holds without assuming~\ref{limitrho-1pexists}.

{\it Remark}.
Recall that $I_0 \equiv 0$ corresponds to  
$p = \frac{1}{6}\,\rho_-^{-\frac{1}{5}}\,\rho^{\frac{6}{5}}$ 
($\rho_- = \mbox{const}$). The associated static perfect fluid
solutions are known explicitly \cite{Chandrasekhar:1957} and read
\begin{equation}\label{exsol} 
v(r) = u(r) = - \frac{M}{\sqrt{\frac{4\pi}{3}\rho_- M^4 + r^2}}\: \qquad (M\geq0).
\end{equation}
Obviously, (\ref{exsol}) are AFMD solutions extending to infinity.

\begin{theorem}\label{J-1theorem} 
Consider an equation of state $\rho(p)$ being ${\mathcal C}^0$, 
piecewise ${\mathcal C}^1$ and monotonic, and assume~\ref{gammaexists}.
\begin{enumerate}
\item[{\bf A}] Let $(v=u-u_S, \rho ,p)$ be an AFMD solution to the Euler-Poisson 
equations~(\ref{poissoneuler}) with $p\leq \tilde{p}$ for some $\tilde{p}>0$.
If $I_{-1} \leq 0$ ($I_{-1} \not\equiv 0$) for all $p\in [0,\tilde{p}]$, 
then the solution has finite extent.
\item[{\bf B}] If $I_{-1} \geq 0$ ($I_{-1} \not\equiv 0$) for all $p\in [0,\tilde{p}]$,
then there is no AFMD solution to the Euler-Poisson 
equations~(\ref{poissoneuler}) satisfying $p\leq \tilde{p}$.
\end{enumerate}
\end{theorem}

{\it Proof.}
The proof is based on Pohozaev's identity in the form encountered in proposition \ref{poho}.
According to the remarks at the end of section \ref{newton:basics}
spherical symmetry is no restriction.
\begin{eqnarray} 
\nonumber
\int\limits_{\mathrm{Ball}(\tilde{r})} d^3x \,(\rho v + 6 p) & = &
-4\pi\int\limits_0^{\tilde{r}} dr r^2 \rho(r) I_0(p(r)) \:=\:
4\pi\int\limits_{p_c}^{p(\tilde{r})} dp r^2 v_{,r}^{-1} I_0 \\ \nonumber
& = &
4\pi\left[ r^2 v_{,r}^{-1} I_{-1} \right]_{p_c}^{p(\tilde{r})} -
4\pi\int\limits_{p_c}^{p(\tilde{r})} dp \frac{d}{dp}(r^2 v_{,r}^{-1}) I_{-1}\\
\label{aI_0integral} 
& = &
4\pi\left[ r^2 v_{,r}^{-1} I_{-1}(p(r)) \right]_{0}^{\tilde{r}} -
4\pi\int\limits_{0}^{\tilde{r}} dr \frac{d}{dr}(r^2 v_{,r}^{-1}) I_{-1}(p(r)) 
\end{eqnarray}
Here, the subscript $c$ refers to the central value of the quantity, e.g., 
$p_c = p|_{r=0}$.
Regularizing and solving the Euler Poisson equations in a small neighborhood of 
$r=0$, we obtain $v(r) = v_c + \frac{4\pi}{6} \rho_c r^2 + o(r^2)$; 
this implies that $r^2 v_{,r}^{-1} I_{-1}(p(r))$
vanishes as $r\rightarrow 0$.
At $r=\tilde{r}$ we have the boundary term 
$\tilde{r}^2 v_{,r}^{-1}|_{\tilde{r}} I_{-1}(p(\tilde{r}))$. 
For solutions with finite extent we may choose $\tilde{r} = R$ in order to
obtain $R^2 v_{,r}^{-1}|_R I_{-1}|_{p = 0} = R^4 M^{-1} I_{-1}|_{p=0} = 0$. 
For AFMD solutions extending to infinity the boundary term vanishes 
as $\tilde{r} \rightarrow \infty\,$: Note that $I_{-1}(p)$ is ${\mathcal C}^1$
at $p=0$ and $I_1|_{p=0} = 0$ as well as $\frac{d}{dp} I_{-1}|_{p=0} = I_0|_{p=0} = 0$. 
Therefore, $I_{-1}(p) = p f(p)$, where $f(p)$ is some function with $f|_{p=0} = 0$. 
Consequently, $\tilde{r}^2 v_{,r}^{-1}|_{\tilde{r}} I_{-1}(p(\tilde{r})) = 
-\tilde{r}^2 \rho(\tilde{r}) \frac{p(\tilde{r})}{p_{,r}(\tilde{r})} f(p(\tilde{r}))\,$, which
must go to zero as $\tilde{r} \rightarrow \infty$, if the AFMD conditions are assumed. 
Now, we define
\begin{equation}
a(r):= r^{-2} \frac{d}{dr}(r^2 v_{,r}^{-1}) = 
4 r^{-1} v_{,r}^{-1} - 4 \pi \rho v_{,r}^{-2} \:.
\end{equation}
The function $a(r)$ is strictly positive, as we will show in the following.
We investigate the function 
$b(r) := r^3 v_{,r}^2 a(r) = 4 r^2 v_{,r} - 4 \pi r^3 \rho\:$.
In a neighborhood of $r=0$, i.e., in the limit $r\rightarrow 0$, $b(r)$ is given by
the positive function
$b_{approx}(r) = 
4 r^2 \frac{4\pi r}{3} \rho_c - 4\pi r^3 \rho_c = \frac{4\pi}{3} r^3 \rho_c \geq 0\:$.
Differentiating $b(r)$ we obtain
$b_{,r}(r) = 4\pi r^2 \rho - 4\pi r^3 \rho_{,r} > 0 \:$, which results from
$\rho_{,r} = \frac{d\rho}{dp} p_{,r} < 0$.
We conclude that $b(r)$ is monotonic and 
$b(r) > 0$ for all $r>0$. Therefore, $a(r) > 0$ for all $r>0$.

Combining (\ref{pohozaevintegrated}) and (\ref{aI_0integral}) 
we obtain
\begin{subequations}
\begin{align}
\int\limits_{\mathbb{R}^3} d^3x \:a(r) I_{-1}(p(x)) &= -\frac{M^2}{R}\:, 
&& \mbox{for solutions with finite extent, and,}
\\[0.1cm]
\int\limits_{\mathbb{R}^3} d^3x \:a(r) I_{-1}(p(x))  &= 0 \:,
&& \mbox{if the fluid extends to infinity,}
\end{align}
\end{subequations}
provided that the solution is AFMD.
From these equations the claim of the theorem follows easily.\proofend

{\it Remark}. (Relationship between theorem~\ref{J0theorem} and theorem~\ref{J-1theorem}).
For the large class of monotonic, piecewise ${\mathcal C}^1$ equations of state
the $I_{-1}$-theorem~\ref{J-1theorem} comprises the $I_0$-theorem~\ref{J0theorem}.
This is simply because $I_{-1}(p)$ is the integral of $I_0(p)$
(compare proposition~\ref{Jinclusions}).

{\it Remark}.
One might conjecture that it is possible to achieve further improvements
of theorem \ref{J-1theorem}
by introducing higher integrals of $I_0(p)$, i.e., $I_{-2}(p)$, $I_{-3}(p)$, etc.
In particular one might think of an extension of theorem~\ref{J-1theorem}  
where $I_{-1}$ is replaced by such an integral.
However, already the $I_{-3}$-analogue of theorem~\ref{J-1theorem} seems to
be wrong (see the following example).
So we may say that theorem~\ref{J-1theorem} is almost the 
``best we can get'', at least in the sense
that probably no further ``integral extension'' exists.

{\it Example}.
Consider the equation of state given by
$\rho(v) = \rho_- v^6 (1 - |v|^{2.5})$, $p(v) = -\int \rho(v) dv$.
This equation of state is differentiable and monotonic (at least up
to $|v_{max}|=0.87$).
Applying the $I_{-1}$-theorem~\ref{J-1theorem} 
we find that $I_{-1}(v)$ is greater than zero up to
$|v_{-1}| = 0.76$. For all solutions with central potential $|v_c|$ less than
$|v_{-1}|$, theorem~\ref{J-1theorem} guarantees infinite extent and infinite mass.
The hypothetical $I_{-3}$-criterion would predict infinite extent
for all $|v_c|$ in the considered range, i.e., for all $|v_c| \leq |v_{max}| = 0.87$.
Now, the numerical observations are the following: 
Every solution corresponding to central potential $|v_c|$ less than a critical
value $|v_{crit}| = 0.86$ has infinite extent and infinite mass.
However, for $|v_c| > |v_{crit}|$ the solution possesses {\it finite} extent!
The numerical results are in accord with theorem~\ref{J-1theorem}, but
they contradict the hypothetical $I_{-3}$-analogue.

\section{Einstein: Basics}
\label{euler-einstein} 

We will now investigate static perfect fluids in General Relativity.
Several quantities and equations which have been 
relevant in the Newtonian description 
possess direct analogues in General Relativity.
To facilitate comparison we will use the same symbols for
corresponding quantities; moreover, wherever permitted, we
will investigate the Newtonian limit.

The metric for a static spacetime can be written as
\begin{equation}\label{fullmetric} 
ds^2 = -V^2(x) dt^2 + g_{ij}(x) dx^i dx^j\:.
\end{equation}
Here, $g$ is a Riemannian metric on the 3-space $M$, 
$V$ a (positive) scalar function on $M$. 
$M$ can be viewed as a hypersurface orthogonal to a timelike
Killing vector whose norm is $V$.

A static perfect fluid solution is a regular solution to the
Euler-Einstein system, i.e., $(V,g_{ij})$ must satisfy
\begin{subequations}\label{einsteineuler}
\begin{align}\label{deltaV}
\Delta V &= 4\pi (\rho+3 p)\,V \\ \label{einstein}
R_{ij} &= V^{-1}\,\nabla_i\nabla_jV + 4\pi (\rho-p)\, g_{ij} \\ \label{eeuler} 
\nabla_i p &= -V^{-1} \,(\rho + p) \nabla_i V \:.
\end{align}
\end{subequations} 
The covariant derivative $\nabla_i$ and the Laplacian $\Delta$ refer to $g$.
Note that (\ref{eeuler}) is not an independent equation; it can be derived
from the Bianchi identity, using (\ref{einstein}) and (\ref{deltaV}).
Equations~(\ref{einsteineuler}) are to be understood in connection
with an equation of state $\rho(p)$ relating the (energy) density and the
pressure.

Integrating (\ref{eeuler}) we see that the function $V$ can be
regarded as a function of $p$.
\begin{equation}\label{Y}
\log Y(p) := \log V(p) - \log V_S = - \int_0^p dp^\prime (\rho(p^\prime) + p^\prime )^{-1} 
=:-\Gamma(p)
\end{equation}
$Y := V/V_S$ is the ``normalized'' quantity. For solutions with finite extent, 
$V_S$ is the value of $V$ at the surface $\{p=0\}$ (such that $Y=1$ there);
at the surface the interior metric~(\ref{fullmetric}) is joined to an exterior vacuum solution
(by standard junction conditions), i.e., 
a Schwarzschild metric in spherical symmetry.
By (\ref{Y}) we see that $\rho$ and $p$ can be viewed as functions of $Y$.
If $\rho(p)$ is monotonic, then $\rho(Y)$ is monotonic as well.
$p(Y)$ and $\rho(Y)$ are related by $-\int_1^Y dY^\prime \rho(Y^\prime) = p Y$.

{\it Remark}.
For smooth, monotonic equations of state, it has been shown \cite{Rendall/Schmidt:1991} that
spherically symmetric solutions of (\ref{einsteineuler}) satisfy
$p(r)\rightarrow 0$, $\rho(r)\rightarrow 0$ and $V(r)\rightarrow 1$ as $r \rightarrow \infty$ 
(where $r$ is an appropriate radial coordinate). 
Hence, $V=Y$, if the fluid extends to infinity.
Like in Newtonian theory $p(r)$ is a decreasing and $V(r)$ is an increasing function;
their central values are again denoted by $p_c$, $V_c$.

{\it Remark}. The Newtonian equations follow from (\ref{fullmetric}) 
and (\ref{einsteineuler}). Writing $V$ as $V = c^2 \exp(\frac{u}{c^2})$ we may regard $u$ as
the generalization of the Newtonian potential. E.g., (\ref{eeuler}) becomes
$\nabla_i p = - (\rho + \frac{p}{c^2}) \nabla_i u$. In the limit $c\rightarrow \infty$
the metric (\ref{fullmetric}) coincides with the Minkowski metric, 
and, e.g., (\ref{eeuler}) becomes $\partial_i p = -\rho \partial_i u$.  
Analogously, in the Newtonian limit, 
the normalized $Y$ is related to the normalized Newtonian potential $v$, 
$Y = V/V_S = \exp(\frac{u-u_S}{c^2}) = \exp(\frac{v}{c^2})$.

\section{Einstein: The quantities $I_i$}
\label{einstein:j0} 

\begin{definition}\label{relGammadef}
Consider a barotropic equation of state $\rho(p)$ ($p \leq p_{max}$) with
$\rho(p)$ piecewise ${\mathcal C}^0$ on $[0, p_{max}]$.
Let $\Gamma(p)$ be defined as
\begin{equation}\label{relGamma}
\Gamma(p) :=  \int_0^p dp^\prime\,(\rho(p^\prime)+ p^\prime)^{-1} \:.
\end{equation}
\end{definition}

\begin{assumption}\label{relgammaexists}
We require that $\Gamma(p)$ exists. Obviously, $\Gamma(p)$ is ${\mathcal C}^0[0,p_{max}]$
\end{assumption}

\begin{assumption}\label{rellimitrho-1pexists}
We assume that the limit $\lim_{p\rightarrow 0} \rho^{-1} p$ exists.
\end{assumption}

\begin{definition}
Consider an equation of state $\rho(p)$ as above.
We define the following quantities:
\begin{subequations}
\begin{align}\label{relJ_0(p)} 
J_0(p) & := \frac{1}{2 \rho}\, [1-e^{-\Gamma}] \,H_0(p) -
3 \rho^{-1} \int_0^p dp^\prime e^{-\Gamma(p^\prime)} 
\frac{1}{\rho(p^\prime)+p^\prime} H_0(p^\prime) \\[0.1cm]
\label{H_0(p)}
& \mbox{where}
\quad\, H_0(p) := \rho(p) + \rho(p) e^{-\Gamma(p)} + 6 p e^{-\Gamma(p)} \\[0.25cm]
\label{relI0} 
I_0(p) & := 1 - e^{-\Gamma(p)} - 6 e^{-\Gamma(p)} \rho^{-1} p \\ \label{relI1}
I_1(p) & := 6 e^{-\Gamma} p \,\rho^{-2}\, \frac{d\rho}{dp} - 5 e^{-\Gamma} (\rho+p)^{-1} \\ 
\label{relI2}
I_2(p) & := 5 (\rho+p) \frac{d\kappa}{dp} + \kappa^2 + 10\kappa \qquad 
(\kappa = \frac{\rho+p}{\rho+3p} \frac{d\rho}{dp}\,)
\end{align}
\end{subequations}
\end{definition}

{\it Remark.}
In complete analogy with section \ref{criteria}, $J_0|_{p=0} = 0$ and $I_0|_{p=0} =0$. 
The comments concerning assumption~\ref{gammaexists} and \ref{limitrho-1pexists}
in section \ref{criteria} apply here as well.


{\it Remark.} (The Newtonian limit).
In the Newtonian limit, i.e., $c\rightarrow \infty$, 
$\Gamma(p) = \int_0^p dp^\prime\,(\rho(p^\prime)+ p^\prime/c^2)^{-1}$ coincides
with the Newtonian $\Gamma(p) = \int_0^p dp^\prime\,\rho^{-1}(p^\prime)$.
According to (\ref{H_0(p)}), $H_0(p) = \rho + \rho \exp(-\Gamma/c^2) + 6 p/c^2 \exp(-\Gamma/c^2)$
becomes $H_0(p) = 2 \rho$ in the limit.
Therefore, $J_0(p) = c^2/(2\rho) [1-e^{-\Gamma/c^2}] \,H_0(p) -
3 \rho^{-1} \int_0^p dp^\prime (\cdots)$ 
approximates $J_0(p) = \Gamma - 6 \rho^{-1} p$, which coincides with
(\ref{J0}). Note that also $I_0$ (\ref{relI0}) coincides with $J_0$ 
in the limit $c\rightarrow \infty$. Analogously, $I_1 \rightarrow J_1$ and
the relativistic $I_2$ (\ref{relI2}) approximates the Newtonian  $I_2$.

\begin{proposition}\label{relJinclusions}
Using the same notation as in proposition \ref{Jinclusions}
we have
\begin{subequations}
\begin{align}
\{I_1 \leq 0\} &\subset \{I_0 \leq 0\} \subset \{J_0 \leq 0\} \\
\{I_1 \equiv 0\} &\equiv \{I_0 \equiv 0\}\equiv \{J_0 \equiv 0\}\\
\{I_1 \geq 0\} &\subset \{I_0 \geq 0\} \subset \{J_0 \geq 0\} \:.
\end{align}
\end{subequations}
\end{proposition}

{\it Proof.}
For the first inclusions note that 
$I_0$ is the integral of $I_1$, i.e., $\frac{d}{dp}\, I_0(p) = I_1(p)\,$, and 
$I_0|_{p=0} = 0$.
Second, with $I_0 \rho = \rho - \rho Y - 6 p Y$ we obtain $H_0 = 2 \rho - I_0 \rho$, whereby
$J_0$ becomes
$J_0 = \frac{1}{2\rho}\,[(1+Y) I_0 \rho - 6 \int_1^Y I_0 \rho dY]$.
Note that $Y\in(0,1]$. Therefore, $I_0 \leq 0$ implies $J_0 \leq 0$
and the other inclusions hold as well. \proofend

{\it Remark}.
$\{J_0\equiv 0\}$ defines the so-called Buchdahl equation of state \cite{Buchdahl:1964}, i.e.,
$p = \frac{1}{6} (\rho_-^{1/5}-\rho^{1/5})^{-1} \, \rho^{6/5}$ ($\rho < \rho_-$). 

{\it Remark}.
The following relations also hold:
$\{I_2 \equiv 0\} \supset \{I_1 \equiv 0\}$ and
$\{I_2 \leq 0\} \subset \{I_1\leq 0\}$. For a proof see \cite{Beig/Simon:1992}.

{\it Comments}.
In the subsequent sections we will only be concerned with the quantities $I_0$ and 
especially $J_0$.
$I_1$ appears in a theorem which basically states 
that a solution (not necessarily AFMD) with $I_1 \leq 0$ must have finite extent.
See~\cite{Simon:2001}.
In General Relativity spherical symmetry of perfect fluid solutions
is a rather involved topic \cite{Lindblom/Masood-ul-Alam:1994, Lindblom:1993}. 
$I_2$ plays the main part in a ``symmetry theorem''. From $I_2 \leq 0$
it can be concluded that the (asymptotically flat) 
solution must necessarily be spherically symmetric \cite{Beig/Simon:1991, Beig/Simon:1992}.

\section{Einstein: Criteria}
\label{einstein:pohozaevandcriterion} 

In analogy to the Newtonian case we formulate criteria for
(in)finiteness of static perfect fluid solutions.
The proofs involve relativistic generalizations of
Pohozaev's identity.

{\it Remark}. The differentiability assumptions for $p(r)$, $V(r)$, etc.\ 
are analogous to the Newtonian case.

\begin{definition}
A solution of (\ref{einsteineuler}) is called AFMD, if for some $\epsilon>0$
\begin{equation}\label{EinsteinAFMD}
1-V = O^\infty(\|x\|^{-1}) \qquad
g_{ij}-\delta_{ij} = O^\infty(\|x\|^{-1}) \qquad
\rho = O^\infty(\|x\|^{-3-\epsilon})
\end{equation} 
in suitable coordinates $\{x^i\}$. By~(\ref{eeuler}), $p= O^\infty(\|x\|^{-4-\epsilon})$.
For a generalization in terms of Sobolev spaces see \cite{Simon:2001}.
\end{definition}

\begin{theorem}\cite{Simon:1993, Simon:2001}.\label{relI0theorem} 
Consider an equation of state $\rho(p)$ being piecewise ${\mathcal C}^0$
and satisfying assumptions~\ref{relgammaexists} and~\ref{rellimitrho-1pexists}. 
\begin{enumerate}
\item[{\bf A}] Let $(Y=V/V_S, \rho ,p)$ be an AFMD solution to the Euler-Einstein 
equations~(\ref{einsteineuler}) with $p\leq \tilde{p}$ for some $\tilde{p}>0$.
If $I_0 \leq 0$ ($I_0 \not\equiv 0$) for all $p\in [0,\tilde{p}]$, 
then the solution has finite extent.
\item[{\bf B}] If $I_0 \geq 0$ ($I_0 \not\equiv 0$) for all $p\in [0,\tilde{p}]$,
then there is no AFMD solution to the Euler-Einstein 
equations~(\ref{einsteineuler}) satisfying $p\leq \tilde{p}$.
\end{enumerate}
\end{theorem}

{\it Sketch of proof}.
The proof of {\bf A} is based on an inequality which can be understood
as some kind of ``Pohozaev inequality''. To establish {\bf B} 
a version of the positive mass theorem is used.
\proofend

{\it Remark}.
For the Buchdahl case $I_0 \equiv 0$, i.e., 
$p = \frac{1}{6} (\rho_-^{1/5}-\rho^{1/5})^{-1} \, \rho^{6/5}$ ($\rho < \rho_-$),
we have the following explicit form for the corresponding
static perfect fluid solutions \cite{Buchdahl:1964}:
\begin{equation}\label{buchdahlsol} 
V(r) = 1 - \frac{M}{\sqrt{\frac{4\pi}{3}\rho_- M^4 + r^2}+\frac{M}{2}}
\qquad g_{ij} = \left(\frac{2}{1+V(r)}\right)^4 \,\delta_{ij} \:.
\end{equation}
Here, $M$ is bounded by $3 M^{-2} < 16\pi \rho_-$.

\begin{definition}
From $Y$, its derivatives, and $p$, $\rho$ we define
the following symmetric tensor $\sigma_{ij}$:
\begin{eqnarray}
\nonumber
\sigma_{ij} & = &\frac{1-Y^2}{Y} \nabla_a\nabla_c Y + 6 Y_{,a} Y_{,c} - 2 g_{ac} Y_{,d} Y^{,d} + 
g_{ac} [-4\pi (1-Y^2) (\rho + p ) - 16\pi p Y (1-Y)] \\ 
& & \quad - 16\pi g_{ij} \,\int_1^Y dY^\prime H_0(Y^\prime)\:, 
\end{eqnarray}
where $H_0(Y)$ is again $H_0(Y) = \rho + \rho Y + 6 p Y\,$.
\end{definition}

\begin{proposition}(Relativistic Pohozaev identity).
Assume spherical symmetry.   
The metric $g_{ij}$ can be written as
$ds^2 = h(r) dr^2 + r^2 d\Omega^2$, and, moreover, there exists 
the conformal Killing vector $\xi_a dx^a= r \sqrt{h} dr$
(``asymptotic dilation''). We have the following identity:
\begin{equation}\label{relPohoagain}
\nabla^i(\sigma_{ij} \xi^j) = -8\pi\,\frac{1}{\sqrt{h}}
\:[(1-Y) H_0(Y) + 6 \int_1^Y dY^\prime H_0(Y^\prime)]
\end{equation} 
\end{proposition}

{\it Proof}. $\nabla_{(a} \xi_{b)} = \frac{1}{\sqrt{h}} g_{ab}$. The tensor 
$\sigma_{ij}$ is divergence free, $\nabla_i \sigma^{ij} = 0$; its
trace is $-8\pi\,[(1-Y) H_0(Y) + 6 \int_1^Y dY^\prime H_0(Y^\prime)]$. \proofend

\begin{proposition}(Integrated version).\label{relpoho}
Consider a spherically symmetric static perfect fluid solution,
i.e., let $V(r)$, $p(r)$, $\rho(r)$ (with $g_{ij}(r)$) be a solution of
(\ref{einsteineuler}). Assume AFMD.
Then, 
\begin{subequations}\label{relpohozaevintegrated}
\begin{align}\label{finiterelpohozaev}
\int\limits_{\mathbb{R}^3} d^3x \,(- \rho \,J_0 ) &= \frac{M^2}{R}\:
\left(1 -\frac{2M}{R}\right)^{-1} \:, 
&& \mbox{for solutions with finite extent, and,}
\\[0.1cm]\label{infiniterelpohozaev} 
\int\limits_{\mathbb{R}^3} d^3x \,(- \rho \,J_0) &= 0 \:,
&& \mbox{if the fluid extends to infinity.}
\end{align}
\end{subequations}
In (\ref{finiterelpohozaev}), $M$ denotes the mass, and $R$ the radius of the finite
fluid object; $\{x^i\}$ are the Cartesian coordinates associated 
with the spherical coordinates $\{r, \theta, \phi\}$, so that $d^3x= r^2 dr d\Omega$.
\end{proposition}

{\it Proof.}
From (\ref{relPohoagain}) we have
$\nabla^i(\sigma_{ij} \xi^j) = -16\pi\frac{1}{\sqrt{h}}\,\rho\,
\,J_0$. We integrate the l.h.\ side of this equation for fluids with
finite extent first.
\begin{equation}\label{pint} 
\int\limits_{\mathrm{Ball}(R)}\sqrt{g}\, d^3x \nabla^i (\sigma_{ij} \xi^j) = 
4\pi R^2 \sigma_{ij}\big|_R\, \xi^j R^{-1} \frac{1}{\sqrt{h(R)}}\, x^i = 
16 \pi R^3 Y_{,r}^2\big|_R\, h(R)^{-1}
\end{equation}
The exterior solution is Schwarzschild, i.e., $V^2(r) = (1-\frac{2M}{r})$ and
$h(r) = V^{-2}(r)$ ($r\geq R$), so that
$Y_{,r}\big|_R = V_S^{-1} V_{,r}\big|_R = V_S^{-2} M R^{-2}$ (with $V_S = V(R)$).
Hence, in (\ref{pint}), $Y_{,r}^2\big|_R\, h(R)^{-1} = M^2 R^{-4} (1-\frac{2M}{R})^{-1}$
and (\ref{finiterelpohozaev}) is established.
For solutions extending to infinity, again
$\int_{\mathrm{Ball}(r)}\sqrt{g} d^3x \nabla^i (\sigma_{ij} \xi^j) = 
4\pi r \sigma_{ij} x^j x^i h(r)^{-1}$. Structurally, 
$\sigma_{ij}$ is mainly built up by terms such as $Y_{,r}^2\, g_{ij}$ or $\rho g_{ij}\,$ 
. Taking account of $g_{ij} x^i x^j = r^2 h(r)$ we obtain
$\sigma_{ij} x^j x^i r h(r)^{-1} \sim
r^3 Y_{,r}^2\,$, or $r^3 \rho$, or the like.
By the AFMD conditions (\ref{EinsteinAFMD}) these terms converge to zero as $r\rightarrow\infty$. 
In similar expressions appearing in $\sigma_{ij}$ it is helpful to use
$h(r) = (1-\frac{2m(r)}{r})^{-1}$ (where $m(r) = 4\pi\int_0^r dr r^2 \rho$),
which follows from the field equations (\ref{einstein}).
Eventually, by (\ref{EinsteinAFMD}), $\sigma_{ij} x^j x^i r h(r)^{-1} \rightarrow 0$ 
($r\rightarrow \infty$). \proofend

\begin{theorem}\label{relJ0theorem} 
Assume spherical symmetry.
Consider an equation of state $\rho(p)$ being piecewise ${\mathcal C}^0$
and satisfying assumptions~\ref{relgammaexists} and~\ref{rellimitrho-1pexists}. 
\begin{enumerate}
\item[{\bf A}] Let $(Y=V_S^{-1} V, \rho ,p)$ be an AFMD solution to the Euler-Einstein 
equations~(\ref{einsteineuler}) with $p\leq \tilde{p}$ for some $\tilde{p}>0$.
If $J_0 \leq 0$ ($J_0 \not\equiv 0$) for all $p\in [0,\tilde{p}]$, 
then the solution has finite extent. 
\item[{\bf B}] If $J_0 \geq 0$ ($J_0 \not\equiv 0$) for all $p\in [0,\tilde{p}]$,
then there is no AFMD solution to the Euler-Einstein 
equations~(\ref{einsteineuler}) satisfying $p\leq \tilde{p}$.
\end{enumerate}
\end{theorem}

{\it Proof.}
The theorem is a direct consequence of 
proposition~\ref{relpoho}.\proofend

{\it Remark}. (Relationship between theorem~\ref{relI0theorem} and theorem~\ref{relJ0theorem}).
The $I_0$-theorem \ref{relI0theorem} 
does not rely on the existence of a conformal Killing vector; spherical
symmetry need not be presupposed.
In spherical symmetry, however, the $I_0$-theorem is covered by the wider 
$J_0$-theorem \ref{relJ0theorem}. See proposition \ref{relJinclusions}.

\section{Examples and Discussion}
\label{examples} 

{\it Example}. (The Generalized Buchdahl family of equations of state).
Consider the following family of equations of state,
\begin{equation}\label{buchdahl}
p(\rho) = \frac{1}{n+1} \,\frac{\rho^{\frac{n+1}{n}}}{\rho_-^{\frac{1}{n}} - \rho^{\frac{1}{n}}}
\qquad (\rho < \rho_-)\:.
\end{equation}
The case $n=5$ corresponds to the Buchdahl equation of state (see section \ref{einstein:j0}).
Equivalently, (\ref{buchdahl}) can be represented by
\begin{equation}
\rho(Y) = \rho_- (1-Y)^n \qquad
p(Y) = \frac{\rho_-}{n+1} \,Y^{-1} (1-Y)^{n+1} \qquad (Y\in (0,1])\:.
\end{equation}
From (\ref{relJ_0(p)}) and (\ref{H_0(p)})
we can compute $J_0(Y)$ which results in
\begin{equation}
J_0(Y) = \left[1-\frac{6}{n+1}\right] \,(1-Y)\, 
\left[1 - \frac{1}{2} (1-Y) (1-\frac{6}{n+2})\right]\:.
\end{equation}
The last bracket is greater than zero for all $Y\in[0,1]$, so we obtain
\begin{equation}\label{buchdahlcases} 
\begin{array}{ccccc}
\bullet & n>5 & \Rightarrow & J_0 > 0 &\\
\bullet & n=5 & \Rightarrow & J_0 \equiv 0 &\\
\bullet & n<5 & \Rightarrow & J_0 < 0 &\:.
\end{array}
\end{equation}
Consequently, theorem \ref{relJ0theorem} applies, i.e., we get finiteness
of the fluid configuration for $n<5$ and infiniteness for $n>5$.
Note incidentally that this result could also have been obtained
using the weaker $I_0$-criterion~\ref{relI0theorem}.

{\it Remark}.
Note that (\ref{buchdahlcases}) also holds
for the polytropes in Newtonian theory. 
We see that in this sense the Generalized Buchdahl family
is the relativistic analogue of the (Newtonian) polytropes.

{\it Example}. (The polytropes in General Relativity).
The polytropic equations of state, $p(\rho) = (n+1)^{-1} \rho_-^{-1/n} \rho^{(n+1)/n}$,
have the following $[\rho(Y), p(Y)]$ form:
\begin{equation}
\rho = \rho_- (n+1)^n (Y^{-\frac{1}{n+1}}-1)^n \qquad
p = (n+1)^n \rho_- (Y^{-\frac{1}{n+1}} -1)^{n+1}
\end{equation}
With some support of a computer algebra program one can show that
\begin{equation}
\begin{array}{ccccc}
\bullet & n>5 & \Rightarrow & J_0 > 0 & (\forall Y)\\
\bullet & n=5 & \Rightarrow & J_0 > 0 & (\forall Y)\\
\bullet & n<5 & \Rightarrow & J_0 < 0 & (Y\geq Y_0(n))
\end{array}
\end{equation}
Here, $Y_0(n)$ is some value of $Y$, where $J_0(Y)$ changes sign;
e.g., $Y_0\big|_{n=4} \approx 0.64$. For $n\rightarrow 5-$, $Y_0(n)$ approximates $1$.
Obviously, for solutions with central ``potential'' $Y_c \geq Y_0$,
theorem \ref{relJ0theorem} applies (predicting finiteness of solutions).

{\it Remark}. Applying the $I_0$-criterion (theorem~\ref{relI0theorem})
to the last example we obtain qualitatively similar results. However, since
it is weaker, the breakdown of the $I_0$-criterion occurs somewhat earlier.
Note further that for certain values $Y_c < Y_0$ one indeed
finds solutions which are not finite
but still AFMD \cite{Nilsson/Uggla:2000}.

{\it Example}. (Asymptotically polytropic equation of state).
In Newtonian theory asymptotically polytropic equations of state
can be written as $\rho(v) = K |v|^n (1 + O(|v|^m))$ (for some $m>0$).
For $m\geq 1$ they coincide with the quasipolytropic equations of 
state \cite{Makino:1998, Simon:2001} (see below).
In analogy to above, up to a certain central density, theorems \ref{J0theorem} and
\ref{J-1theorem}
ensure finiteness/infiniteness (for $n<5$/$n>5$ respectively).
Again, for certain values exceeding the critical central density,
one finds (AFMD) solutions extending to infinity or
finite solutions (for $n<5$ or $n>5$ respectively).
Compare partly with the remarks at the end of section \ref{newton:pohozaevandcriteria}.
Note that 
these examples indicate that the breakdown of the $I_0$- or $I_{-1}$-criterion 
is not at all artificial;
on the contrary it seems to be an important feature, 
anticipating the appearance of the counterintuitive behavior we observe for higher densities.

{\it Example}. (Degenerate matter).
An equation of state satisfying 
$p = K \rho^{(n+1)/n} ( 1 + O(\rho^{1/n}))$ (as $\rho\rightarrow 0$) 
is called quasipolytropic \cite{Simon:2001}.
Note the following important result:
Spherically symmetric perfect fluid solutions corresponding to quasipolytropic equations
of state with $n<3$ have finite extent \cite{Makino:1998}. 
Degenerate matter is usually described by quasipolytropic
equations of state with $n=3/2$, so we know that the
corresponding star model is finite.
In this connection it is instructive to investigate how restrictive 
the $J_0$-criterion is.
Consider the equation of state of a completely degenerate, ideal Fermi gas
(see, e.g., \cite{Shapiro/Teukolsky:1983}). 
For degenerate electrons (white dwarfs) $J_0$ is negative for all
admissible densities (and far beyond), i.e., as long as the energy density is
dominated by the rest mass of the ions.
For neutron stars (degenerate neutrons) $J_0 \leq 0$ is valid up to 
densities of $10^{17} \mbox{g}\, \mbox{cm}^{-3}$ which is beyond the top end of neutron
star densities.
As an example for a more realistic neutron star model
take the Harrison-Wheeler equation of state (see, e.g., \cite{Shapiro/Teukolsky:1983} for
an introduction). $J_0 \leq 0$ holds for $\rho \leq 7\times 10^{11} \mbox{g}\,\mbox{cm}^{-3}$. 
This is slightly beyond the density region where ``neutron drip'' occurs. Sometimes
the Harrison-Wheeler equation of state is replaced by different equations of state for
such densities, so the outcome is satisfying enough also in this case.
Summing up we see that the $J_0$-theorem \ref{relJ0theorem}
is rather generally applicable also for these (already known) situations.

\section{Pohozaev Identities}
\label{methodofconstruction} 

The present section provides the necessary background material to understand
how ``Pohozaev-like'' identities come into existence:
We outline a method of constructing
such identities.
In particular, we re-derive Pohozaev's identity \cite{Pohozaev:1965} in the Newtonian case 
(see equation (\ref{pohozaev})) and we 
construct its direct analogue for the relativistic case (see (\ref{relPohoagain})).

{\it Basics}.
Let $(M,g)$ be a Riemannian manifold, $\mathrm{dim} M = 3$, with
Ricci tensor $R_{ij}$ and curvature scalar $R$. 
Assume that $M$ possesses a conformal Killing vector $\xi$, 
$\nabla_{(a} \xi_{b)} = \frac{1}{3} \nabla^c \xi_c g_{ab}$.
Consider the conformal rescaling
\begin{equation}
\tilde{g}_{ij} = \Omega^2\: g_{ij}\:.
\end{equation}
We define 
\begin{equation}\label{tautilde}
\tilde{\tau}_{ij} := \tilde{R}_{ij} - \frac{1}{2}\,\tilde{g}_{ij}\,\tilde{R}\:,
\qquad \tau_{ij} := \Omega \,\tilde{\tau}_{ij}\:.
\end{equation}
The tensor $\tilde{\tau}$ is divergence free (with respect to $\tilde{\nabla}$) and
symmetric, thus
$\nabla_i \tau^{ij} = \Omega^{-1} \, \tau^k_k\, g^{ij} \nabla_i \Omega$,
wherefore
\begin{equation}\label{tau}
\nabla_i(\tau^{ij} \xi_j) = \tau^k_k \:
\left[ \frac{1}{3} (\nabla^l\xi_l) + \Omega^{-1} g^{ij} (\nabla_i\Omega) \:\xi_j \right]\:.
\end{equation}

{\it Aim}.
Our aim is to construct
Pohozaev-like identities, i.e., relations of the form
\begin{equation}\label{poho-like}
\nabla^i(\alpha_i(x)) = \beta(x)\, J + \gamma(x)\:,
\end{equation}
where $\alpha_i(x)$ shows sensible
asymptotic behavior, and $\beta(x)$ and $\gamma(x)$ are
strictly positive (or negative) functions. We require $J$ to consist of
expressions manifestly determined by the equation of state, such as
$\rho$, $p$, $\frac{d\rho}{dp}$, $\Gamma(p)$, and others.
Such identities might give rise to nontrivial theorems; compare with the
Pohozaev identity (\ref{pohozaev}) and the resulting theorem \ref{J0theorem}.
Our starting point to obtain these identities will be equation (\ref{tau}).

{\it Identities in the Newtonian case}.
We will illustrate the procedure for the (easier) Newtonian case first. 
The relativistic case is completely analogous, even though slightly complex.
Recall that Newtonian perfect fluid solutions are described by the 
potential $v$, a scalar field on a flat background $g_{ij} = \delta_{ij}$.
We consider conformal rescalings where the conformal factor $\Omega$ is a
function of $v$, i.e., $\Omega = \Omega(v)$.
\begin{equation}\label{Omegarescaling} 
\tilde{g}_{ij} = \Omega^2(v)\: g_{ij}\qquad g_{ij} =\delta_{ij}\:.
\end{equation}
From (\ref{tautilde}) we get the following for the tensor $\tau_{ij}$ 
(we use the notation $\Omega^\prime = \frac{d\Omega}{dv}$):
\begin{equation}\label{explicittau} 
\tau_{ij} = (2 \Omega^{-1} \Omega^{\prime 2} - \Omega^{\prime\prime}) v_{,i} v_{,j} -
\Omega^\prime v_{,ij} + 
(\Omega^{\prime\prime} - \Omega^{-1} \Omega^{\prime 2}) v^{,k} v_{,k} \delta_{ij} +
\Omega^\prime \Delta v \delta_{ij} \:.
\end{equation}
On $(\mathbb{R}^3, \delta_{ij})$ we may use the dilation $\xi_i = x_i$ 
as the conformal Killing vector, i.e.,
$\partial_{(a} x_{b)} = \delta_{ab}$, and (\ref{tau}) becomes
\begin{equation}\label{newtontau} 
\partial_i(\tau^{ij} x_j) = \tau^k_k + 
\Omega^{-1} \,\tau^k_k \,x^i \Omega_{,i} \:. 
\end{equation}

Based on (\ref{newtontau}) the idea is to modify the tensor $\tau_{ij}$ in order to 
get a r.h.\ side of the form (\ref{poho-like}).
We define
\begin{subequations}\label{sigmad}
\begin{align}\label{sigma}
\sigma_{ij} &:= \tau_{ij} + d_{ij} \\ \label{d} 
d_{ij} &:= d_1 \delta_{ij} + d_2 \frac{1}{r^2} x_{(i} v_{,j)} + 
d_3 v_{,i} v_{,j}
\end{align}
\end{subequations}
In (\ref{d}) the $d_i$ are functions $d_i(v)$, $\xi_i = x_i$ is the conformal
Killing vector, $r$ its norm. Hence, $d_{ij}$ is a tensor consisting of
functions of $v$, derivatives of $v$ and the conformal Killing vector.
The ansatz (\ref{d}) for $d_{ij}$ is general provided that we assume
spherical symmetry. This is because we are only interested in the trace
of $d_{ij}$ and the expression $x^j \partial^i d_{ij}$ (see below).
Terms like $v_{ij}$ 
, $v_{,k} v^{,k} \delta_{ij}$, or $x_i x_j$ can be
subsumed under the ansatz~(\ref{d}).

Replacing $\tau_{ij}$ by $\sigma_{ij}$ in (\ref{newtontau}) we obtain
$\partial_i(\sigma^{ij} x_j) = \tau^k_k + 
\Omega^{-1} \,\tau^k_k \,x^i \Omega_{,i} + d^k_k + x^j \partial^i d_{ij}$,
i.e., explicitly,
\begin{eqnarray}\nonumber
\partial_i(\sigma^{ij} x_j) & = & 
8 \pi \rho \Omega^\prime(v) + 3 d_1(v) + 4 \pi \rho d_2(v) \,+  \\
\label{identities} 
& & + \,v_{,r}^2\, [\hat{\Omega} + d_2^\prime - d_3] 
+ r v_{,r}\, [ 8 \pi \rho \Omega^{-1} \Omega^{\prime 2} + d_1^\prime + 8\pi \rho d_3 ] + \\
\nonumber 
& & +\, r v_{,r}^3\, [ \hat{\Omega} \Omega^{-1} \Omega^\prime + d_3^\prime ]\:.
\end{eqnarray}
Here, $\hat{\Omega}$ abbreviates 
$\hat{\Omega} = 2 \Omega^{\prime\prime} - \Omega^{-1} \Omega^{\prime 2}$.
Without loss of generality we have restricted ourselves to spherical symmetry.

For a large class of $\Omega(v)$ and the free functions $d_i(v)$, 
the r.h.\ side of equation (\ref{identities}) is exactly of the form (\ref{poho-like}).
Integrating the total divergence on the l.h.\ side we obtain
\begin{equation}\label{boundaryterm} 
\frac{1}{4\pi}\int\limits_{\mathrm{Ball}(r)} d^3x \partial^i(\sigma_{ij} x^j) =
r^3 v_{,r}^2 \Omega^{-1} \Omega^{\prime 2} + 2 r^2 v_{,r} \Omega^\prime +
r^3 \,[d_1 + \frac{v_{,r}}{r} d_2 + v_{,r}^2 d_3]\:.
\end{equation}
If this expression can be evaluated at $r=R$ for finite fluid solutions,
then (\ref{identities}) is a candidate for a sensible Pohozaev-like identity.
For AFMD solutions the existence of the limit $r\rightarrow\infty$
of (\ref{boundaryterm}) should be investigated.

{\it Example}. (Pohozaev identity). We choose $\Omega(v) = v^2$. 
Note that $\hat{\Omega} = 0$. 
Choose $d_2 \equiv 0$ and $d_3 \equiv 0$, so (\ref{identities}) becomes
\begin{equation}
\partial_i(\sigma^{ij} x_j) = 16 \pi \rho v  + 3 d_1(v) 
+ r v_{,r}\, [ 32 \pi \rho + d_1^\prime ] \:.
\end{equation}
The last bracket vanishes if $d_1(v) = 32 \pi p$, so that we obtain
\begin{equation}
\partial_i(\sigma^{ij} x_j) = 16 \pi (\rho v  + 6 p)\:,
\end{equation}
which coincides with the Pohozaev identity (\ref{pohozaev}).
Moreover, as expected, $\sigma_{ij}$ (as given by (\ref{sigmad}) which is 
essentially (\ref{explicittau}))
coincides with (\ref{sigmaij}).

{\it Remark}.
An even wider class of identities could be constructed admitting
more general conformal factors in (\ref{Omegarescaling}), e.g.,
let $\Omega = \Omega(v,v_{,r})$.
In this way one could possibly obtain identities adapted to
specific problems (i.e., classes of equations of state).


{\it Relativistic Pohozaev identity}.
The relativistic case can be treated in analogy to the Newtonian one. However, in order to
simplify the presentation we confine ourselves to a certain choice of the
conformal factor $\Omega$ from the beginning.
The metric for a spherically symmetric 3-space can be written as
\begin{equation}
g_{ij} dx^i dx^j = h(r) dr^2 + r^2 (d\theta^2 + \sin^2\theta d\phi^2) \qquad (h > 0) \:.
\end{equation}
Such a metric is conformally flat and possesses the conformal Killing vector $\xi^a$,
\begin{equation}
\xi_a dx^a = r \sqrt{h} dr \:,\qquad \nabla_{(a} \xi_{b)} = \frac{1}{\sqrt{h}} g_{ab}\:.
\end{equation}
For perfect fluid solutions $h(r) = (1-\frac{2m(r)}{r})^{-1}$, where
$m(r) = 4\pi\int_0^r dr r^2 \rho$, so that --for AFMD solutions-- 
$\xi$ can be viewed as an ``asymptotic dilation''.

We make the following choice for the conformal factor $\Omega$:
\begin{equation}\label{Omegachoice} 
\tilde{g}_{ij} = \Omega^2(Y)\: g_{ij}\qquad \Omega(Y) = (1-Y)^2
\end{equation}
From~(\ref{tautilde}) we calculate the tensor $\tau_{ac}$,
\begin{eqnarray}
\nonumber
\tau_{ac} & = & \frac{1-Y^2}{Y} \nabla_a\nabla_c Y + 6 Y_{,a} Y_{,c} - 2 g_{ac} Y_{,d} Y^{,d} + \\
\label{reltau} 
& & \quad + g_{ac} [-4\pi (1-Y^2) (\rho + p ) - 16\pi p Y (1-Y)] 
\end{eqnarray}
With $\sigma_{ij} := \tau_{ij} + d_{ij}$ and the simplified ansatz 
$d_{ij} = d_1(Y) \,g_{ij}$ we modify equation (\ref{tau}) in order to get
\begin{subequations}
\begin{align}
\nabla_i(\sigma^{ij} \xi_j) & = \frac{1}{\sqrt{h}}\, \tau^k_k + 
\Omega^{-1} \,\tau^k_k \,g^{ij} \xi_i \Omega_{,j} + 
\frac{1}{\sqrt{h}} d^k_k + \xi^j \nabla^i d_{ij} \\ \label{interim} 
& = \frac{1}{\sqrt{h}} (\tau^k_k + d^k_k) + h^{-1} Y_{,r} \xi_r \,
[ \frac{d}{dY} d_1(Y) + 16\pi \rho + 16\pi \rho Y + 96 \pi p Y ] \:.
\end{align}
\end{subequations}
Equation (\ref{interim}) suggests the choice 
$\frac{d}{dY}\,d_1(Y) = -16\pi H_0(Y) = - 16\pi \rho - 16\pi \rho Y - 96 \pi p Y$.  
Consequently, $d_1(Y) = -16 \pi \int dY H_0(Y)$, and
\begin{equation}\label{relPohozaev}
\nabla_i(\sigma^{ij} \xi_j) = -8\pi\,\frac{1}{\sqrt{h}}
\:\left[(1-Y) H_0(Y) + 6 \int\limits_1^Y dY^\prime H_0(Y^\prime)\right]
\end{equation} 
\begin{equation}
\mbox{with} \qquad H_0(Y) = \rho + \rho Y + 6 p Y\:.
\end{equation}
Equation (\ref{relPohozaev}) is the relativistic
counterpart of Pohozaev's identity (see (\ref{relPohoagain})).
Recall that the tensor $\sigma_{ij}$ is given by $\sigma_{ij} = \tau_{ij} + d_{ij}$, i.e.,
\begin{equation}
\sigma_{ij} = \mbox{(\ref{reltau})} - 16\pi g_{ij} \,\int_1^Y dY^\prime H_0(Y^\prime)\:.
\end{equation}

{\it Remark}. In order to simplify the presentation we 
have chosen a particular conformal factor in (\ref{Omegachoice}).
Obviously, without specifying $\Omega(Y)$, via a more general ansatz for $d_{ij}$
(as in \ref{d}), we would arrive at relativistic analogues of (\ref{identities}).
Since we have not used such identities in the paper, we refrain from presenting
the (lengthy) formulae here.

\begin{appendix}
\section{Mathematical properties of $\rho(p)$}
\label{appendixA}

Assumptions~\ref{gammaexists} and~\ref{limitrho-1pexists} in section \ref{criteria} 
are independent.
From the existence of $\Gamma(p)$, by~(\ref{gammaandlimit}), 
we may conclude that $\rho \,p^{-1}$ cannot be bounded as $p \rightarrow 0$.
However, $\lim_{p\rightarrow 0} \rho^{-1} p$ need not exist as is shown
by the following example. Define a ${\mathcal C}^\infty$ function $s(p)$ with 
the properties $0 \leq s \leq 1$, $s(p_n) = 1$, 
$\mathrm{supp}\, s \subseteq \bigcup_n [p_n - p_n^2, p_n + p_n^2]$; where
$p_n = 2^{-n}$ ($n\in\mathbb{N}$). 
Consider the equation of state $\rho(p) = p\,(s(p) + p^\epsilon)^{-1}$ ($0<\epsilon<1$).
$\rho$ is in ${\mathcal C}^\infty(0,1)$ and in ${\mathcal C}^0[0,1)$, with $\rho(p=0) = 0$.
Furthermore, $\Gamma(p)$ exists, as can easily be seen. However, the limit
$\rho^{-1} p$ as $p\rightarrow 0$ does not exist.
Conversely, take the equation of state $\rho(p) = - p \log p$ ($p\leq p_{max}< 1$). The limit
$\lim_{p\rightarrow 0} \rho^{-1} p$ exists. However, 
$\int_\epsilon^p dp^\prime p^{\prime\:-1} (\log p^\prime)^{-1} = 
\log(-\log p)-\log(-\log\epsilon)$ 
diverges as $\epsilon\rightarrow 0$, so that $\Gamma(p)$ does not exist.

If the equation of state $\rho(p)$ is monotonic on $[0,\epsilon)$, then
assumption~\ref{limitrho-1pexists} follows from assumption~\ref{gammaexists}.
To show this note that
\begin{equation}
\Gamma(p) = \int_0^p dp^\prime \rho^{-1}(p^\prime) \geq \rho^{-1}(p) \int_0^p dp^\prime
= \rho^{-1}(p)\, p\:.
\end{equation}
This is because
$\rho^{-1}(p^\prime) \geq \rho^{-1}(p)$ $\forall p^\prime \leq p$. 
Letting $p\rightarrow 0$ the claim is established.

Evidently, if \ref{gammaexists} 
and~\ref{limitrho-1pexists} hold, $\lim_{p\rightarrow 0}\rho^{-1} p = 0$.
For $\rho|_{p=0} > 0$ this is obvious, so let us indirectly assume that 
$\rho|_{p=0} = 0$ with $\lim_{p\rightarrow 0} \rho^{-1} p = c \neq 0$.
Choose $\tilde{p} > 0$,
such that $\rho^{-1} p > c/2 \quad \forall p < \tilde{p}$.
\begin{equation}\label{gammaandlimit}
\infty > \int_0^{\tilde{p}} \rho^{-1}(p) dp =
\int_0^{\tilde{p}} \rho^{-1}(p) p \,p^{-1} dp > 
\frac{c}{2} \,\int_0^{\tilde{p}} p^{-1} dp = \infty
\end{equation} 
A contradiction.

If it exists, $\lim_{p\rightarrow 0} (\frac{d\rho}{dp})^{-1} = 0$;
this is shown by simply applying de l'Hospital's rule.
Note, however, that there are monotonic equations of state satisfying 
assumption~\ref{gammaexists} (and, consequently, assumption~\ref{limitrho-1pexists}),
whose limit $\lim_{p\rightarrow 0} \frac{dp}{d\rho}$ is not defined.
As an example, take $\rho(p) = \int_0^p p^{-1/(n+1)} (1-s(p))\, dp$
(where $s(p)$ is defined as above).
\end{appendix}

\subsection*{Acknowledgements}

I am grateful to R.\ Beig for helpful discussions and his support
and to C.\ Uggla for inspiring conversations. 
In particular I would like to thank W.\ Simon who initiated the 
work on finiteness of static perfect fluids;
his comments were always very valuable and highly appreciated.
This work was supported by the {\it Austrian Academy of Sciences}
(Doktorandenprogramm der \"Osterreichischen Akademie der Wissenschaften).



\end{document}